# Shooting High or Low: Do Countries Benefit from Entering Unrelated Activities?

Flávio L. Pinheiro[1,*,†], Aamena Alshamsi[2,1], Dominik Hartmann[3,4], Ron Boschma[5,6], and César A. Hidalgo[1,*,†]

[1] Collective Learning Group, The MIT Media Lab, Massachusetts Institute of Technology, Cambridge MA
[2] Masdar Institute of Science and Technology, Abu Dhabi, UAE
[3] Fraunhofer Center for International Management and Knowledge Economy, Leipzig, Germany
[4] Chair of Innovation Management and Innovation Economics, University of Leipzig, Germany
[5] Department of Human Geography and Planning, Utrecht University, the Netherlands
[6] UiS Business School, Stavanger Centre for Innovation Research, University of Stavanger, Norway

* Correspondence to: flaviopp@mit.edu, and hidalgo@mit.edu
† Current Address: The MIT Media Lab, 75 Amherst Street, Cambridge, MA 02139

**Abstract**

Countries tend to diversify their exports by entering products that are related to their current exports. Yet this average behavior is not representative of every diversification path. In this paper, we introduce a method to identify periods when countries enter unrelated products. We analyze the economic diversification paths of 93 countries between 1965 and 2014 and find that countries enter unrelated products in only about 7.2% of all observations. We find that countries enter more unrelated products when they are at an intermediate level of economic development, and when they have higher levels of human capital. Finally, we ask whether countries entering more unrelated products grow faster than those entering only related products. The data shows that countries that enter more unrelated activities experience a small but significant increase in future economic growth, compared to countries with a similar level of income, human capital, capital stock per worker, and economic complexity.

# Introduction

A well-established fact in the literature on international and regional economic development is that countries (Hidalgo, Klinger, Barabási, & Hausmann, 2007), regions (Coniglio, Lagravinese, Vurchio, & Armenise, 2018; Gao et al., 2017; Neffke, Henning, & Boschma, 2011; Zhu, He, & Zhou, 2017) and cities (Boschma, Balland, & Kogler, 2015), are more likely to enter economic activities that are similar to the ones they already have. This *principle of relatedness* has been observed for activities as diverse as products (Hidalgo et al., 2007), industries (Neffke et al., 2011), technologies (Boschma et al., 2014) and research areas (Guevara, Hartmann, Aristarán, Mendoza, & Hidalgo, 2016). Yet, what is true on average is not true for every instance. While countries and regions are more likely to enter related economic activities, sometimes they deviate from this behavior and enter unrelated activities. Repeatedly, scholars have stated that unrelated diversification is needed to avoid lock-in and to ensure long-term economic development (Saviotti & Frenken, 2008), but systematic evidence, documenting episodes of unrelated diversification and exploring its potential benefits, is yet lacking (Boschma, 2017).

In this paper, we introduce a method to identify when countries enter unrelated activities and use it to identify the countries that do so more frequently. Finally, we study the macroeconomic consequences of such deviations by studying whether countries that entered more unrelated activities grew faster.

But how common it is for a country to enter an unrelated activity? In a dataset containing 93 countries during the period 1965-2014, we find that in 92.8% of the observed cases countries enter the export of products that are more related than the average product in their option set. This means that in only 7.2% of the cases countries deviate from the principle of relatedness and enter products that are on average, less related than their option sets. But do these deviations happen at any level of development? Or are they more frequent for developing or developed countries? The data shows that countries enter relatively more unrelated activities when they are at an intermediate level of economic development, and also when they have a relatively high level of human capital. This is true when we measure a country's level of development using GDP per capita, export diversity, or economic complexity. In fact, the level of relatedness of new economic activities follows a U-shaped curve, with countries at both low and high levels of development entering primarily related activities, and countries with an intermediate level of economic development entering more unrelated activities. To the best of our knowledge, this is the first time that the stage of development of a country is linked with the type of related activity development.

Finally, we ask whether countries entering more unrelated activities grow faster, after controlling for their initial level of GDP per capita, economic complexity, and human capital. We find that countries entering more unrelated activities experience a small but significant increase in their subsequent growth performance of about 0.5% per annum. In other words, our findings show that deviations from the principle of relatedness occur especially in countries at an intermediate stage of development, and that countries that succeed in developing unrelated activities experience a small but significant boost in economic growth.

## Literature Review

Knowledge production is often conceived as a process of recombining existing ideas (Romer, 1994; Weitzman, 1998). To compete, organizations explore different parts of the knowledge space (Breschi, Lissoni, & Malerba, 2003), searching for new knowledge while being constrained by the limits of bounded rationality (Simon, 1972). Since organizations have limited access to information, and do not have a perfect capacity to absorb, process, and respond to new information (Cohen & Levinthal, 1990), they experience a cost of diversification that decreases with the level of relatedness of activities (Atkinson & Stiglitz, 1969; Chatterjee & Wernerfelt, 1991).

Countries and regions are collections of organizations limited by tacit and embedded knowledge (Gertler, 2003), and hence, they should also experience strong path dependencies in their diversification processes (Frenken & Boschma, 2007; Hidalgo et al., 2007; Neffke et al., 2011). Instead of emerging randomly, new activities build on and combine existing local capabilities, resulting in distinctive technological and industrial profiles of countries and regions (Rigby & Essletzbichler, 1997).

A large volume of studies provides strong evidence supporting the notion that diversification in countries and regions is path dependent (Boschma, 2017; Hidalgo et al., 2017). *Hidalgo et al.* (2007) showed how countries expand their mix of exports around the products in which they already established a comparative advantage. *Neffke et al.* (2011) used information on product portfolios of plants to show that regions tend to diversify into new industries related to existing local industries. *Kogler et al.* (2013), *Rigby* (2015), *Boschma et al.* (2015) and Petralia et al. (2017) among others, used measures of technological relatedness between patent classes to show that countries and cities develop new technologies related to existing local technologies.

But relatedness is not the only factor shaping the path dependencies of economies. A key driving force behind the distinctiveness of regional trajectories is the complexity of knowledge (Kogut & Zander, 2003). *Fleming and Sorenson* (2001) define the complexity of a technology in terms of the number of components and the interdependence between those components. *Hidalgo and Hausmann* (2009) derive a formula for the complexity — or knowledge intensity — of products and places by defining knowledge intense places as those that produce knowledge intense activities, and knowledge intense activities as those produced by knowledge intense places. This circular definition is mathematically solvable using the linear algebra concept of eigenvectors. Yet, the more colloquial definition of knowledge complexity is still tied to the idea that complexity is reflected in a wide range of capabilities. These complex products tend to be produced by relatively few knowledge intense countries, and hence, can support higher wages for the workers employed in these industries (Hartmann, Guevara, Jara-Figueroa, Aristarán, & Hidalgo, 2017; Hausmann et al., 2014; Hidalgo, 2015; Hidalgo & Hausmann, 2009).

Recently, these measures of complexity have also been extended to the production of technology. *Balland and Rigby* (2016) found huge variations in the complexity of knowledge produced across U.S. cities (i.e. few metropolitan areas produce the most complex technologies, while many cities produce the least complex ones) which also correlates highly with the long-run economic performance of cities.

So, in principle, it should be beneficial for a country to build comparative advantages in complex technologies. Yet, for many countries, this is difficult to achieve, because accumulating these capabilities is particularly difficult when the capabilities needed are unrelated to the ones available in a location. So, a key goal for the literature on diversification of territories is to understand the ability of countries or regions to defy the principle of relatedness and enter relatively unrelated and sophisticated economic activities. Yet, we know little about unrelated diversification.

One recent line of research, exploring unrelated diversification, has looked at regional variation, asking whether unrelated diversification prevails in certain countries and regions. *Xiao et al.* (2018) showed, for instance, that European regions with a higher innovation capacity are more inclined to enter less related industries. *Boschma and Capone* (2015a) observe that Western European economies also tend to diversify more into unrelated industries than Eastern European economies, which follow their existing industries more closely. Another line of research focuses on which agents are responsible for more unrelated diversification. For instance, *Neffke, Hartog, Boschma and Henning* (2018) show that external agents (entrepreneurs, firms) coming from outside the region, are more likely to introduce unrelated diversification and to shift specializations of regions. This is especially true for new subsidiaries that are established by large firms located in other regions because subsidiaries can still rely on internal resources of the parent organization that are unavailable in their host region (Crescenzi, Gagliardi, & Iammarino, 2015). *Boschma and Capone* (2015a) explore the role of institutions in unrelated diversification, showing that countries with more liberal, and less coordinated forms of capitalism, are more likely to diversify into more unrelated activities. *Montresor and Quatraro* (2017) found that regions with a strong presence of key enabling technologies had a tendency to diversify into more unrelated technologies. And *Petralia et al.* (2017) showed that high-income countries have a higher tendency to diversify into more unrelated and sophisticated technologies.

But despite all of this recent work, we still know little about when countries are more likely to enter unrelated activities, and about the potential economic benefits of unrelated diversification (Boschma, 2017). Here, we develop a method that attempts to fill that gap by analyzing the relative degree of relatedness of a new economic activity, and investigating whether countries that succeed in entering more unrelated activities also experience higher economic growth, after controlling for their initial GDP per capita, economic complexity, capital stock per worker, and human capital.

## Materials and Methods

### Data

We use international trade data from the MIT's Observatory of Economic Complexity (atlas.media.mit.edu). We use the SITC-4 rev 2 product classification, since it provides the longest time series: 1962 to 2014. This dataset combines exports data from 1962 to 2000, compiled by *Feenstra et al.* (2005), and data from the U.N. Comtrade for the period between 2001 and 2014. This data contains detailed trade information for 225 countries and 1,004 distinct products.

We reduce noise coming from underreporting and from variations in the size of the economies of countries and products by using several filters. First, we filter out all city-sized national economies, by discarding all countries with a population of fewer than 1.2 million citizens and a total trade below USD 1 billion in 2008. Moreover, economies for which no reliable data was available—such as Iraq (IRQ), Chad (TCD), Macau (MAC) and Afghanistan (AFG)—were discarded. Finally, we employ several time-dependent filters. All yearly trade flows valued at less than 5,000 USD were discarded. Also, we discarded all products whose exports value is equal to zero for more than 80% of the countries. Additionally, products with a global export of less than USD 10 million and countries whose exports equal to zero for 95% of the products are also excluded. These filters allow us to remove countries and products that do not have a significant relevance in the global trade and that would, otherwise, introduce noise to the analysis. After applying these filters, our final sample consists of 117 countries, representing 97.45% of global GDP and 86.67% of global trade in 2008. We use 2008 as a point of reference since this is the last year for which we can estimate regressions for new products (from 2008 to 2010) and validate them during the following four years (from 2010 to 2014).

We use GDP, population and human capital from the Penn World Tables (PWT 9.0). The GDP data comes from real GDP National Accounts, measuring GDP in constant USD in 2005 (Feenstra, Inklaar, & Timmer, 2015). To measure changes in a country's productive structure—which require long time series data—we restrict our analysis to a subset of 93 countries for which data is available during the entire interval of analysis, from 1965 until 2010 (see Appendix A). These countries correspond to 86.73% of global GDP and 73.31% of global trade in 2010.

## The Economic Complexity Index

The Economic Complexity Index (ECI) is a measure of the knowledge intensity of economies and products that can be computed from trade data. Here, we compute the Economic Complexity Index (ECI) and the Product Complexity Index (PCI) following (Hidalgo et al., 2007; Hidalgo & Hausmann, 2009). To compute ECI, we define $R_{cp}$ as a matrix of the Revealed Comparative Advantages (RCA) connecting countries to their significant exports (i.e. the products they export more than what we

*Pinheiro, Alshamsi, Hartmann, Boschma and Hidalgo (2018)*

expect based on a country's total exports and a product's global market). Formally $R_{cp}$ is defined as

$$R_{cp} = \left(\frac{X_{cp}}{\sum_{cp'} X_{cp'}}\right) \Big/ \left(\frac{\sum_{c'p} X_{c'p}}{\sum_{c'p'} X_{c'p'}}\right) \qquad (1)$$

where $X_{cp}$ is a matrix summarizing the dollar exports of country $c$ in product $p$. Then, we define $M_{cp} = 1$ if a country has a comparative advantage in a product ($R_{cp} \geq 1$) and $M_{cp} = 0$ otherwise. $M_{cp}$ contains information about a country's significant exports. Using $M_{cp}$ we define the diversity of a country as the number of products that it exports ($k_c = \sum_p M_{cp}$) with revealed comparative advantage, and the ubiquity of a product as the number of countries that export such product with revealed comparative advantage ($k_p = \sum_c M_{cp}$).

Moreover, we define the product basket of a country (*c*) at year (*y*) as all products for which that country holds RCA greater or equal to one. The remaining products, which have RCA lower than 1, comprise the option set of a country (*c*) at year (*y*). This option set is denoted by $O_{cy}$.

Using the above definitions, the knowledge intensity of a country (ECI) is computed as the average knowledge intensity of the products it exports (i.e. the average PCI of its exports). Conversely, we compute the knowledge intensity of a product (PCI) as the average knowledge intensity of the countries exporting it. This circular argument gives rise to the following iterative mapping:

$$\text{ECI}_c = \frac{1}{k_c} \sum_p M_{cp} \text{PCI}_p \tag{2a}$$

$$\text{PCI}_p = \frac{1}{k_p} \sum_c M_{cp} \text{ECI}_c \tag{2b}$$

Putting (2b) into (2a) provides an eigenvalue equation whose solution is a country's economic complexity index.

$$\text{ECI}_c = \sum_p \frac{M_{cp}}{k_p k_c} \sum_c M_{cp} \text{ECI}_c \tag{3}$$

ECI offers a measure of the knowledge in an economy. PCI is a measure of the knowledge in an industry. Like its counterpart, it can be computed by solving the following eigenvalue equation:

$$\text{PCI}_p = \sum_c \frac{M_{cp}}{k_p k_c} \sum_p M_{cp} \text{PCI}_p \tag{4}$$

### Relatedness and Proximity in The Product Space

We estimate the proximity between two products by looking at the probability they are co-exported. Formally, the proximity between products *p* and *p'* ($\phi_{pp'}$) is the minimum of the conditional probability that a country has a Revealed Comparative Advantage (RCA) in both products:

$$\phi_{pp'} = \frac{\sum_c M_{cp} M_{cp'}}{\max(k_p, k_{p'})} \quad (5)$$

For instance, a proximity of 0.4 between two products means that there is at least a 40% chance that a country with RCA in one product has RCA in both products (Hidalgo et al., 2007).

We then use this proximity to estimate the relatedness between the products that a country exports and each of the products it does not export. The resulting quantity is commonly referred to as the density, $\omega_{cp}$, of product $p$ in country $c$ and computed as:

$$\omega_{cp} = \frac{\sum_{p'} M_{cp'} \phi_{pp'}}{\sum_{p'} \phi_{pp'}} \quad (6)$$

Higher density products are products that are more related/similar to the export capacities of a country, whereas lower density products correspond to unrelated/farther away products.

### Identifying New Products

Here we propose a new methodology to quantify the relative level of relatedness and complexity of newly developed products by a country, which requires identifying when a country enters a new product. To that end, we say that a country (*c*) enters a new product (*p*), between years *y* and *y'*, when it is able to jump from RCA lower than 1.0 at year *y* to RCA above or equal to 1.0 at year *y'*.

However, since the RCA depends on export data, the temporal patterns tend to be very noisy. As a result, this methodology, alone, is prone to false-positives (*e.g.*, temporal spikes in the exports of a country that do not repeat). To correct for *false-positives*, we implement two additional conditions: First, we consider a backward condition that requires that a country *c* had an RCA lower than 1.0 over product *p* for four consecutive years before *y*; Secondly, a forward condition requires that a country *c* is able to maintain a RCA above or equal to 1.00 over product *p* during the following 4 years after *y'*. We only consider as newly developed products those that verify both conditions.

## Results

Following the relatedness literature (Boschma, 2017; Hidalgo et al., 2017), we test firstly if, on average, countries are more likely to diversify towards products that are related to their current exports. Figure 1a shows a visualization of the product space (the network connecting similar products $\phi_{pp'}$), highlighting, as an example, the products for which South Korea had RCA in 1987. Products that are related to Korea's exports in that year (products with high $\omega_{cp}$) are those products that are connected to many of South Korea's current exports.

Figures 1b and 1c reproduce the basic finding of the relatedness literature. They show that the probability that a country will develop RCA in a product is larger for more related products. Figure 1b estimates relatedness using the density measure described in equation 5. Density estimates the relatedness between all of the products exported by a country and each potential new export product. Figure 1c estimates relatedness using the proximity of each potential product to the closest exported product. Both methods show that countries are — on average — more likely to start exporting products that are related to their current exports. These results are robust to different threshold values of RCA from 0.5 to 1.5.

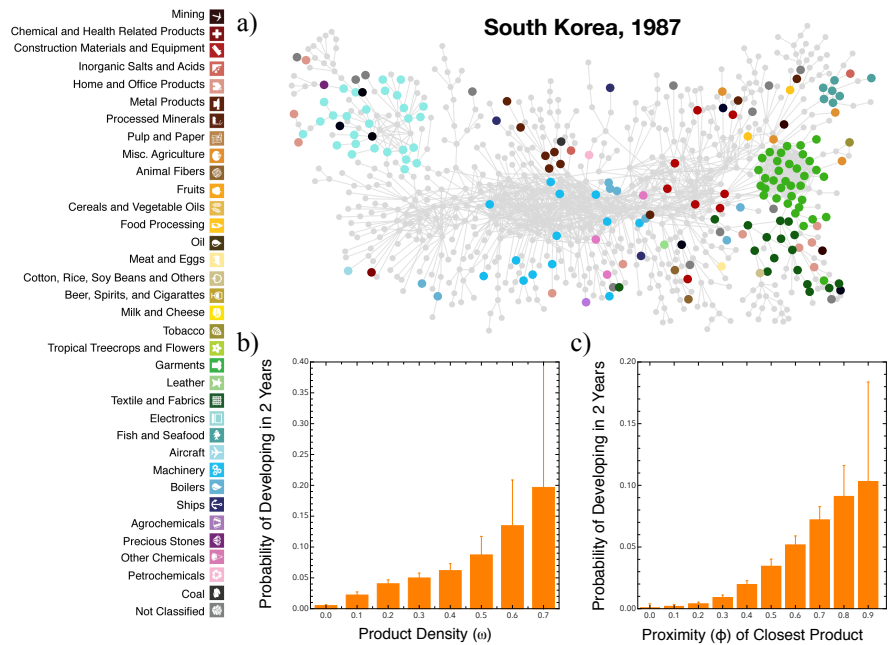

**Figure 1** – Product Space and the Principle of Relatedness. Panel a) shows the Product Space, highlighting, as an example, the products in which South Korea had RCA over 1 in 1987. The colors of the nodes correspond to the sectors described in the legend on the left. Panel b) shows the probability of developing a new export product over a four years period as a function of density. Panel c) shows the probability of developing a new export product over a four years period as a function of the proximity to the closest developed product. A new export product is considered to be developed if in a four-year period it undergoes a transition from $R_{cp} < 1$ to $R_{cp} \geq 1$.

Yet, both of these measures of relatedness cannot be readily used for comparisons across countries. Density estimates the distance between a product and all of the related products, and hence, has the problem that it does not fall within the same range of values for all countries. For instance, a diversified economy, like that of Sweden or Switzerland, will have relatively high values of density for most products, whereas a less diversified economy, like that of Angola or Ecuador for instance, will have relatively low values of density for most products. Thus, the correlation between density and entering related economic activities observed in Figure 1b could be driven by more diversified countries entering more activities. Looking at the proximity to the closest product partly corrects for this, but introduces another problem, since having one close product does not mean that the country has all of the related knowledge needed to develop an activity.

To avoid these limitations, we introduce a variation of this measure we call *Relative Density* ($\widetilde{\omega}_{p,c,y}$). Unlike density ($\omega_{p,c,y}$), which is an absolute measure, *Relative Density* compares the density of a country's new exports with that country's *option set*[1]. We define the diversification option set of a country as the set of all products it does not export. Hence, the relative density of product *p*, in year *y* is computed as:

$$\widetilde{\omega}_{p,c,y} = \frac{\omega_{p,c,y} - \langle \omega_{c,y} \rangle_O}{\sigma_O(\omega_{c,y})} \quad (7)$$

where $\langle \omega_{c,y} \rangle_O$ is the average density of all products in $O_{c,y}$, and $\sigma_O(\omega_{c,y})$ is the standard deviation of the density of the same group of products. Hence, *relative density* ($\widetilde{\omega}_{p,c,y}$) compares the relatedness of a country's new exports to that of all potential new exports, and hence, is comparable across countries because it tells whether a country enters a product that was more, or less, related than the average product in its option set.

Likewise, we define the Relative Complexity of a product ($\widetilde{PCI}_{p,c,y}$) as the complexity of products compared to the average complexity (PCI) of all potential new exports of a country. The relative complexity of a product *p*, in relation to the option set of country *c* in year *y* can be computed as:

---

[1] Alternatively, we could have made the measure relative by ranking products according to their level of relatedness, but we chose not to use rankings because rankings are uniformly distributed.

$$\widetilde{\text{PCI}}_{p,c,y} = \frac{\text{PCI}_{p,c,y,} - \langle PCI_{c,y} \rangle_O}{\sigma_O(PCI_{c,y})} \tag{8}$$

where $\langle PCI_{c,y} \rangle_O$ is the average PCI of all products in $O_{c,y}$, and $\sigma_O(PCI_{c,y})$ is the standard deviation of the PCI of the same set of products. Hence, $\widetilde{\text{PCI}}_{p,c,y}$ is a standardization of the complexities of products in relation to the option set. So, this measure of *Relative Complexity* tells us whether a country entered a product that was more or less complex than the average product the country was not exporting.

Figure 2a and 2b illustrate these ideas by comparing the diversification option sets and the products entered by South Korea in 1994-1996 and by Chile in 1989-1991. Since both *Relative Density* and *Relative Complexity* are measured with respect to a country's option set, the products that are not exported by these countries (shown in grey), are distributed around the center (0,0), which represents the average complexity and density of a country's option set. We highlight South Korea and Chile's new exports in red. We consider that country *c* became an exporter of product *p* when the revealed comparative advantage (RCA) increased from $R_{cp} < 1.00$ for five consecutive years, to $R_{cp} \geq 1.00$ after four years. We also require RCA to remain above 1.00 for the next four years to ensure the country sustained a relatively high level of comparative advantage after the transition. We notice these results are robust to using different RCA cutoffs. Figure 2a shows that most of the products that South Korea entered in 1994-1996 had a low *Relative Density* and a high *Relative Complexity*. This means South Korea entered products that were more unrelated and sophisticated than the average options it had available. Conversely, Figure 2b shows that between 1989 and 1991 Chile entered products that were less sophisticated and more related than the average product in Chile's option set.

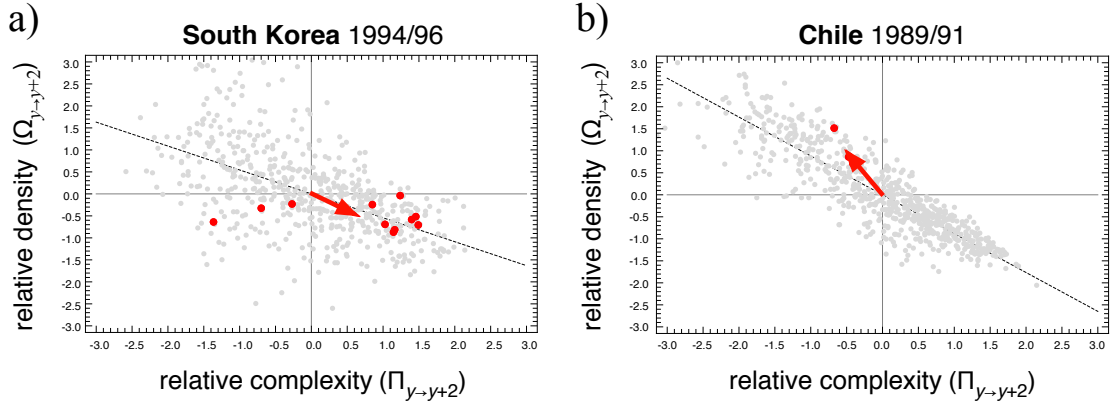

Figure 2 - a) New export products developed by South Korea from 1995 to 1997; the red arrow indicates the respective development direction. b) New export products developed by Chile from 1989 to 1991. From 1995 to 1996 South Korea moved towards unrelated and complex activities. Chile instead moved from 1989 to 1991 towards related and simple activities.

We can characterize the average "direction" of a country's new exports using a two-dimensional vector (Figure 2a and 2b) pointing to all the new products that an economy began exporting in a given time window. In the case of South Korea in 1994-1996, this vector points to the bottom right (relatively more complex and more unrelated). In the case of Chile in 1989-1991, this vector points to the top left (relatively less complex and unrelated). Formally, we define this vector as:

$$\mathbf{J}_{c,y\to y'} = \{\Pi_{c,y\to y'}, \Omega_{c,y\to y'}\} \qquad (9)$$

where $\Omega_{c,y\to y'}$ corresponds to the simple average of $\widetilde{\omega}_{cp}$ and $\Pi_{c,y\to y'}$ to the average of $\widetilde{PCI}_{cp}$ of the set of products developed from $y$ to $y'$.

Figure 3a and 3b show the distribution for the two components of this vector ($\Omega_{c,y\to y'}$ and $\Pi_{c,y\to y'}$) for all considered countries in the dataset. The distribution of relative relatedness ($\Omega_{c,y\to y'}$) shows mostly positive values, confirming the principle of relatedness. Relative relatedness is positive in 92.8% of cases, meaning that in most cases countries enter products that are more related than the average product in their option sets. Instead, unrelated diversification is rare: countries enter unrelated products in only 7.2% of all observations.

The distribution of relative complexity $\Pi_{c,y\to y'}$, is almost centered, and only slightly biased towards negative values, meaning that countries are likely to enter products of a complexity level that is similar to the average complexity of the products in their option set. Put together, these two distributions tell us that countries, on average, shoot for related products with a sophistication that is slightly lower than the average in their option set.

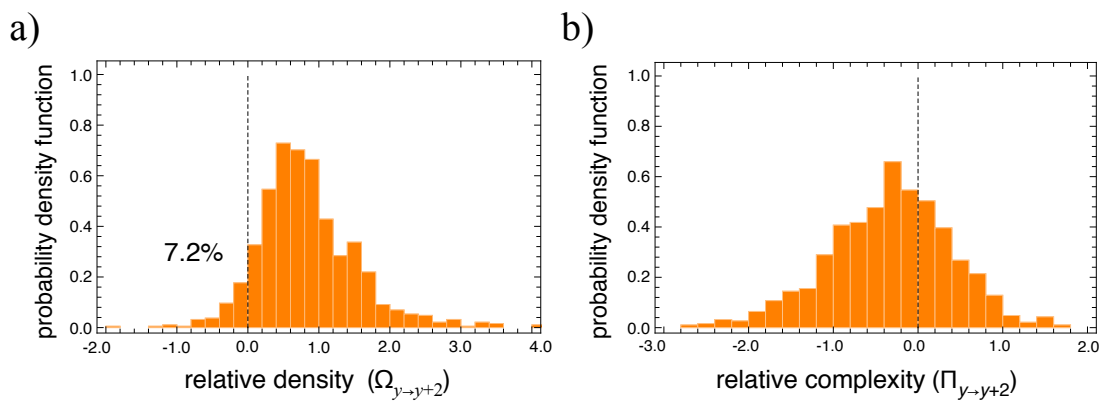

Figure 3 - The distributions of the development directions in respect to the average relatedness (panel a) and the average complexity (panel b) of newly developed export products between 1970 and 2010, in non-overlapping intervals of 2 years.

Figure 4a summarizes the development direction vector by showing the association between the average change in relatedness and complexity for all countries between 1970 and 2010. Here, we consider products entering in a 2-year time window, but the results are robust to considering other time windows. Figure 4b shows the same plot but aggregates the data by country. Both figures reveal a negative correlation between relative density and complexity, meaning that when countries enter more unrelated activities (deviating more from relatedness), they also enter more sophisticated (higher complexity) activities. Figure 4c shows how the slope of the aggregated behavior by country (slope of the linear regression shown in Figure 4b) changed between 1970 and 2010 in 2-year intervals. It shows that while the trend remains negative—*i.e.*, countries that develop more complex varieties tend to develop towards more unrelated activities—this behavior has become less accentuated over the years.

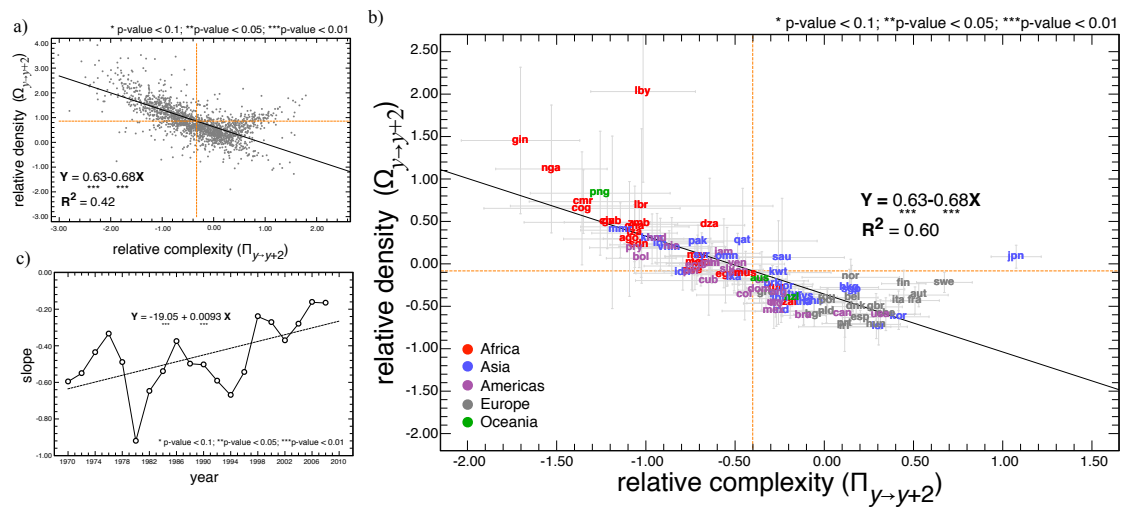

Figure 4 - Development directions of all 93 countries between the years 1970 and 2010. In panel a) each dot corresponds to the average development direction of a country in a year, for all 2 years non-overlapping intervals between 1970 and 2010, the black line shows the best linear model over the sample of 1,747 observations. Panel b) shows the development directions aggregated by country during the entire interval, the black line shows the best linear model. Dots are colored according to the continent of the respective country. In both panels a) and b), the orange dashed lines indicate the average values of the relative density $\Omega_{c,y \to y+2}$ (horizontal) and the relative complexity $\Pi_{c,y \to y+2}$ (vertical). Panel c) shows the slope in panel b) when data is aggregated in 2-year intervals between 1970 and 2010.

Next, we explore whether countries are more likely to enter more unrelated activities at particular stages of their economic development. Figure 5a-b shows the relative relatedness of new products entered by a country ($\Omega_{c,y \to y\prime}$) as a function of their level of economic complexity and GDP per capita. In both cases, we find a minimum level of relative relatedness—meaning that countries are most likely to enter more unrelated activities—at an intermediate stage of development. This stage is at an economic complexity of about 1.01 and a GDP per capita of about USD 25k. Similarly, we look at the relative complexity of new exports ($\Pi_{c,y \to y\prime}$) as a function of the economic complexity and GDP per capita of countries (Figure 5c-d). In both cases, we find that countries with higher levels of development, measured in terms of economic complexity and GDP per capita, enter higher complexity products, as expected.

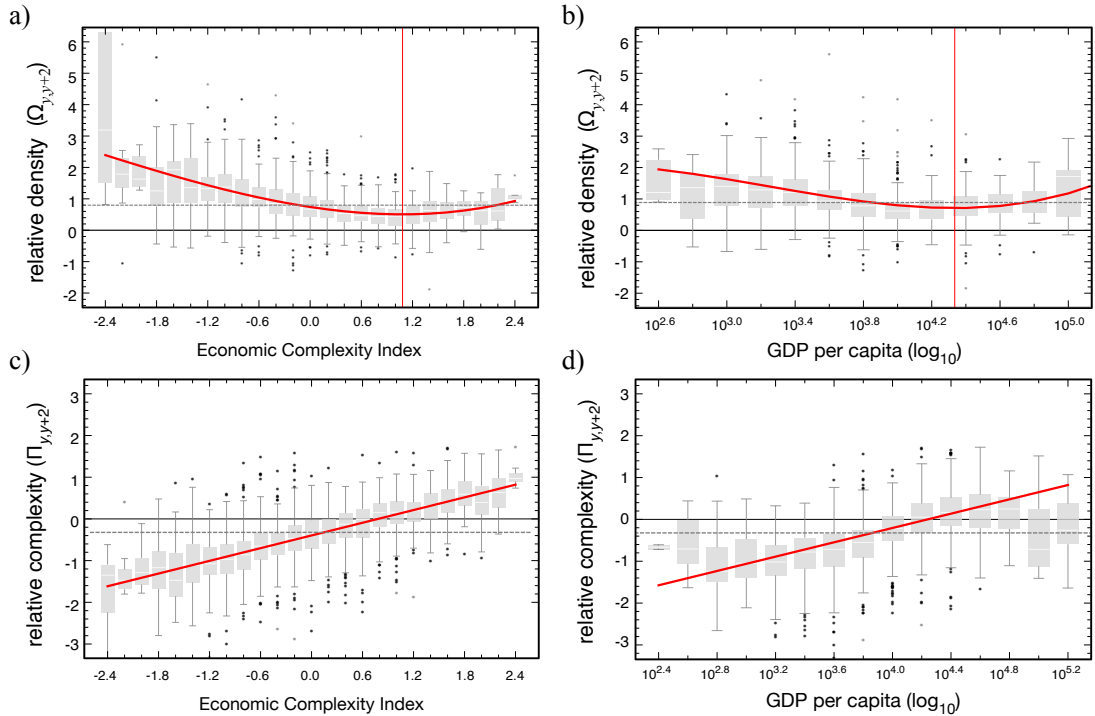

Figure 5 - Top panels show the deviation in the relatedness of newly developed products ($\Omega_{c,y \to y+2}$) as a function of the Economic Complexity Index (a) and GDP chained PPP *per capita* in 2011 USD (b). Vertical red line shows the minimum of the best fit third order polynomial. Horizontal dashed line shows the average $\Omega_{c,y \to y+2}$ over all countries, which is $\langle \Omega_{c,y \to y+2} \rangle = 0.77$. Bottom panels show the deviation in the complexity of newly developed products ($\Pi_{c,y \to y+2}$) as a function of the Economic Complexity Index (c) and GDP chained PPP *per capita* in 2011 USD (d). Red lines show the best linear fit to the data. Horizontal dashed line shows the average $\Pi_{c,y \to y+2}$ over all countries, which is $\langle \Pi_{c,y \to y+2} \rangle = -0.35$.

Next, we explore which factors influence the development of unrelated varieties. To that end, we model the *relative relatedness* of new products ($\Omega_{c,y \to y\prime}$) as a function of the initial level of GDP *per capita*, human capital[2], capital stock[3] per worker, and economic complexity of a country. We also consider the effect of quadratic terms in the regression, as Figures 5a and 5b suggest the existence of a non-linear relationship between a country's level of economic complexity and the number of new unrelated activities it enters.

Table 1 – Models regressing the average Relative Density of newly developed Products ($\Omega_{c,y \to y+2}$) as a function of linear and quadratic terms of the Economic Complexity Index (ECI), Log of GDP per capita (GDP), Initial Human Capital (HC) and Initial Capital Stock per worker (Stock) (linear only).

|  | Average Relative Density of newly developed Products ($\Omega_{c,y \to y+2}$) | | | | |
| --- | --- | --- | --- | --- | --- |
|  | (1) | (2) | (3) | (4) | (5) |
| ECI | -0.362*** |  | -0.296*** | -0.291*** | -0.296*** |
|  | (0.018) |  | (0.026) | (0.027) | (0.026) |
| ECI² | 0.155*** |  | 0.161*** | 0.161*** | 0.152*** |
|  | (0.015) |  | (0.015) | (0.015) | (0.016) |
| Initial ln GDP$_{pc}$ |  | -0.695*** |  |  |  |
|  |  | (0.191) |  |  |  |
| Initial (ln GDP$_{pc}$)² |  | 0.028*** |  |  |  |
|  |  | (0.011) |  |  |  |
| Initial Human Capital |  |  | -0.140*** | -0.128*** | -0.401** |
|  |  |  | (0.039) | (0.042) | (0.179) |
| Initial (Human Capital)² |  |  |  |  | 0.058 |
|  |  |  |  |  | (0.039) |
| Initial ln Capital Stock |  |  |  | -0.013 |  |
|  |  |  |  | (0.017) |  |
| Constant | 0.721*** | 4.689*** | 0.964*** | 1.081*** | 1.234*** |
|  | (0.073) | (0.848) | (0.100) | (0.180) | (0.207) |
| Observations | 1,511 | 1,511 | 1,511 | 1,511 | 1,511 |

---

[2] Human Capital was obtained from PWT v9 table (Feenstra et al., 2015). This quantity estimates educational attainment across countries and is measured by using the average years of schooling from *Barro and Lee* (2013) and an assumed rate of return to education (Psacharopoulos, 1994).

[3] Capital Stock of each country divided by the number of employed people, obtained from the PWT v9 table (Feenstra et al., 2015).

|  | | | | | |
|---|---|---|---|---|---|
| R² | 0.225 | 0.114 | 0.231 | 0.231 | 0.232 |
| Adjusted R² | 0.214 | 0.101 | 0.220 | 0.220 | 0.220 |
| F Statistic | 20.547*** | 9.101*** | 20.338*** | 19.475*** | 19.567*** |

Notes: *p<0.1; **p<0.05; ***p<0.01
All models control for years fixed effects
Observations collected between 1970 and 2010 in two-year steps

Table 1 summarizes the results of five specifications of this model. The model confirms the observation of the quadratic relationships, meaning that countries are more likely to develop unrelated activities when they are at an intermediate stage of development and when they have relatively high levels of human capital. We also find that Economic Complexity captures this relationship much more accurately than GDP per capita ($R^2 = 22\%$ vs. 11%), and human capital appears to have an impact even when controlling for the level of economic complexity of a country (Model 5).

Table 2 – Models regressing the GDP per capita annualized growth as a function of country-specific features and the relative complexity ($\Pi_{c,y \to y+2}$) and relatedness ($\Omega_{c,y \to y+2}$) of newly developed products to the averages of the option set of each country over two-year time interval. Note that a negative coefficient associated with $\Omega_{c,y \to y+2}$ implies, by definition, that countries see a faster GDP per capita growth when developing unrelated activities.

|  | GDP per capita annualized growth ($g_{c,y \to y+2}$) | | | | | | |
|---|---|---|---|---|---|---|---|
|  | (1) | (2) | (3) | (4) | (5) | (6) | (7) |
| $\Omega_{c,y \to y+2}$ | -0.005*** |  | -0.005*** |  | -0.005*** |  | -0.005*** |
|  | (0.001) |  | (0.002) |  | (0.001) |  | (0.002) |
| $\Pi_{c,y \to y+2}$ |  | 0.003** | -0.00003 |  |  | 0.003* | -0.001 |
|  |  | (0.001) | (0.002) |  |  | (0.002) | (0.002) |
| Initial ECI |  |  |  | 0.006*** | 0.005*** | 0.004** | 0.005*** |
|  |  |  |  | (0.002) | (0.002) | (0.002) | (0.002) |
| Initial ln GDP$_{pc}$ |  |  |  | -0.004* | -0.005* | -0.005** | -0.005* |
|  |  |  |  | (0.002) | (0.002) | (0.002) | (0.002) |
| Initial ln Pop |  |  |  | 0.002** | 0.001* | 0.001** | 0.001* |
|  |  |  |  | (0.001) | (0.001) | (0.001) | (0.001) |
| Initial Human Capital |  |  |  | 0.007*** | 0.007*** | 0.007*** | 0.007*** |
|  |  |  |  | (0.003) | (0.003) | (0.003) | (0.003) |
| Initial ln Capital Stock |  |  |  | -0.004** | -0.004** | -0.004** | -0.004** |
|  |  |  |  | (0.002) | (0.002) | (0.002) | (0.002) |
| Constant | 0.041*** | 0.038*** | 0.041*** | 0.101*** | 0.109*** | 0.106*** | 0.108*** |
|  | (0.004) | (0.004) | (0.004) | (0.012) | (0.012) | (0.013) | (0.013) |
| Observations | 1,511 | 1,511 | 1,511 | 1,511 | 1,511 | 1,511 | 1,511 |
| R² | 0.079 | 0.073 | 0.079 | 0.115 | 0.122 | 0.117 | 0.122 |

| Adjusted R² | 0.067 | 0.061 | 0.066 | 0.101 | 0.107 | 0.102 | 0.107 |
| F Statistic | 6.386*** | 5.871*** | 6.078*** | 8.071*** | 8.267*** | 7.872*** | 7.946*** |

Notes: *p<0.1; **p<0.05; ***p<0.01
All models control for year fixed effect
Observations collected between 1970 and 2008 in two-year steps

Finally, we explore whether developing more unrelated products correlates with the future economic growth of countries. We model annual growth in GDP per capita as a function of the *relative relatedness* of new products ($\Omega_{c,y \to y'}$), the *relative complexity* of new products ($\Pi_{c,y \to y'}$), and the initial level of GDP per capita, human capital, capital stock per worker, and economic complexity of a country. Formally, we use a model of the form:

$$g_{c,y \to y'} = \alpha_1 \times \Omega_{c,y \to y'} + \alpha_2 \times \Pi_{c,y \to y'} + \boldsymbol{\beta}_{c,y} \times \mathbf{X}_{c,y} + \mu_y + \epsilon \qquad (10)$$

where the dependent variable $g_{c,y \to y'}$ represents the annualized GDP *per capita* growth from year $y$ to $y'$, $\mathbf{X}_{c,y}$ is a vector of country-specific features (human capital, capita; stock per worker, Economic Complexity, and GDP per capita) measured at year $y$, $\boldsymbol{\beta}_{c,y}$ is the vector of coefficients associated with the country-specific features, and $\mu_y$ controls for year fixed effects. By the definition of $\Omega_{c,y \to y+2}$ (measuring relatedness), a negative $\alpha_1$ coefficient implies that deviations towards more unrelated products would contribute to positive growth of GDP *per capita*. We test five different models to explain yearly GDP per capita growth of newly developed products in a 2-year interval ($y' = y + 2$). Table 2 summarizes the results of different models.

Model (1) and Model (2) show, respectively, how the average deviations in relatedness ($\Omega_{c,y\to y+2}$) and the complexity ($\Pi_{c,y\to y+2}$) of the newly developed products predict GDP *per capita* growth independently. We find that the coefficient associated with $\Omega_{c,y\to y+2}$ is statistically significant. One standard deviation towards more unrelated products contributes with an increase in 0.5% in the yearly growth of GDP *per capita*. Likewise, one standard deviation towards more complex products contributes with 0.3%. The interaction term between $\Omega_{c,y\to y+2}$ and $\Pi_{c,y\to y+2}$ is not statistically significant (see Table B2 in Appendix B). Model (4) shows how country-specific features alone - such as initial Population Size, Economical Complexity Index, initial GDP *per capita*, initial Human Capital Index and initial Capital per worker - explain the yearly growth of GDP *per capita*. Model (5) and Model (6) test the statistical significance of $\Omega_{c,y\to y+2}$ and $\Pi_{c,y\to y+2}$ when controlling for the country-specific features. In these models, only the term capturing deviations in relatedness remains significant, which implies that the deviations in complexity of newly developed products are fully explained by traditional metrics. In sum, models (1) to (6) showed that countries benefit while entering unrelated activities. These results remain valid for extended time periods (see effects on economic growth in subsequent 3 to 10 years in Appendix B).

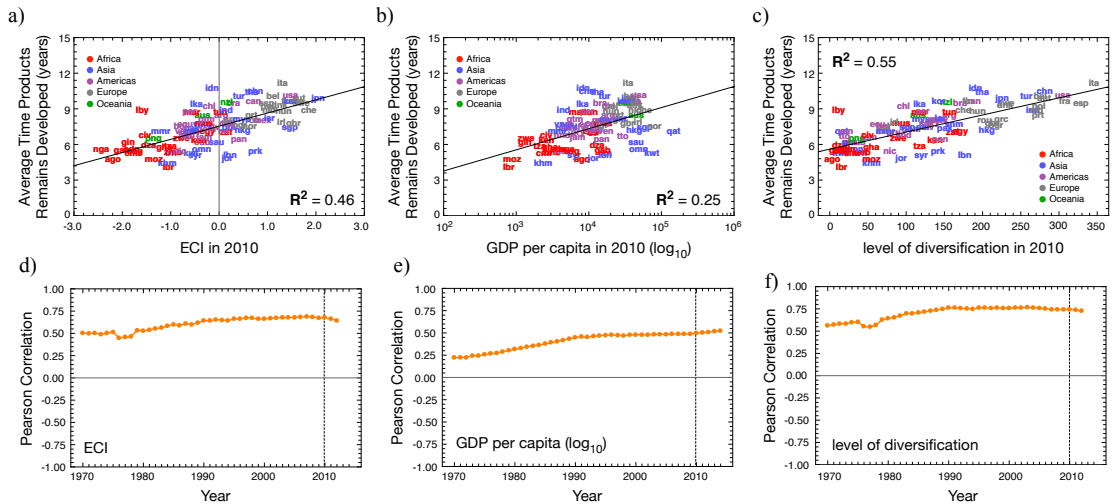

Figure 6 - Average time a new product remains in the product basket of a country (in years) as a function of the Economic Complexity Index (a), GDP per capita (b), level of diversification (c), and average density (relatedness) of the option set (d). The level of diversification is estimated by the number of products in the product basket. Only products that were developed between 1970 and 2010 are considered. Panels d, e and f shows how the Pearson correlation coefficient changes when the ECI, GDP per capita and level of diversification are taken from different reference years is shown, dashed vertical lines show the reference values used in panels a, b and c.

Finally, we analyze whether the ability of a country to sustain new products is dependent on its stage of development. Indeed, countries with higher levels of ECI, GDP per capita, mean density of the option set, and diversification (measured in 2010) were able to keep newly developed products for longer time intervals in their product baskets (see Figure 6a-c). We find a positive (and even slightly increasing) association between the stage of development and the ability to sustain new products for all years between 1970 and 2010 (see Figure 6d-f).

## Discussion

Understanding economic diversification is a long-standing puzzle. There is growing consensus in the literature that diversification is a path dependent process. Yet, deviations from these path dependencies, albeit infrequent, are expected to have an important development role, since they may be key to overcome the industrial lock-in affecting the world's less diversified countries. Here, we explored the frequency of unrelated diversification, the factors correlating with it, and its economic implications. To do this, we introduced the concepts of relative density and complexity, which allowed us to compare jumps in the product space across countries, since these are measures that are relative to each country's option set. We found that unrelated diversification tends to happen at an intermediary stage of economic development, which is evidenced by the U-shape relationship between the relative density of newly developed products and the level of Economic Complexity and GDP per capita. Also, countries with higher human capital tend to develop towards unrelated activities, even after controlling for Economic Complexity. Finally, we showed that unrelated diversification is correlated with higher growth rates. However, these results do not imply that all countries are able and should invest in unrelated variety growth. Indeed, related variety growth remains to be a key driver of economic development (Boschma & Capone, 2015b; Hidalgo & Hausmann, 2009) over most part of the economic development and diversification process of countries. Nonetheless, there is also a critical intermediary stage of economic development in which countries are able and thus can chose to take risks in investing into unrelated variety growth. But why do countries diversify to unrelated varieties at this intermediate stage?

One possible explanation is the shifting shape of the option set of a country. At low levels of development, countries tend to be close to products that are of a low level of complexity, meaning that they see an option set with a negative correlation between relatedness and complexity (i.e. the most unrelated products are the most complex). At high levels of development, countries are confronted with an opposite option set: higher complexity products are among the most related. So, the low hanging fruits of developing countries include mostly products of low complexity, whereas the low hanging fruits of developed countries include mostly sophisticated, complex products. This implies that there must be a transition somewhere in between.

A measure that can capture this relationship is the correlation between relatedness and complexity for the products in a country's option set. Formally, this is given by:

$$\rho_{c,y} = corr(\widehat{PCI}_{p,c,y}, \widetilde{\omega}_{p,c,y}) \tag{11a}$$

$$\rho_{c,y} = \frac{\sum_{O_c}(\widehat{PCI}_{p,c,y} - \langle\widehat{PCI}_{p,c,y}\rangle)(\widetilde{\omega}_{p,c,y} - \langle\widetilde{\omega}_{p,c,y}\rangle)}{\sum_{O_c}(\widehat{PCI}_{p,c,y} - \langle\widehat{PCI}_{p,c,y}\rangle)^2(\widetilde{\omega}_{p,c,y} - \langle\widetilde{\omega}_{p,c,y}\rangle)^2} \tag{11b}$$

where equation (11b) is estimated over the option set ($O_c$), that is the set of all the products that at year $y$ country $c$ has RCA $< 1.0$. What does this quantity translate into? A negative correlation ($\rho_{c,y} < 0$) indicates that a country is closer to the least complex products, a positive correlation ($\rho_{c,y} > 0$) implies that countries are closer to the most complex products available to develop, and a null correlation ($\rho_{c,y} = 0$) means that country is close to both simple and complex products.

Figure 7 shows the relationship between relative density (Y-Axis) and relative complexity (X-Axis) of the option set of South Korea for the years 1987 (a), 2002 (b) and 2010 (c). The red line in each panel shows the best linear fit. Hence in 1987, South Korea exhibited a negative correlation ($\rho_{c,y} < 0$), and thus was closer to the least complex products. Conversely, in 2010 South Korea showed a positive correlation ($\rho_{c,y} > 0$) and was closer to the most complex products in the option set, so it did not need to develop unrelated varieties any longer. In the intermediate stage, in 2002, South Korea was closer to both complex and simple varieties, showing a null correlation. Accordingly, the economy of South Korea underwent a transition from 'being close to simple products' to a situation where it is 'closer to complex products'. Moreover, on the right side of panels (d), (e) and (f) the distribution of the product occupation is depicted for countries that exhibit the different correlation levels exemplified in (a), (b) and (c). Countries closer to the least complex products tend to occupy the peripheral region of the Product Space (d), in contrast, countries closer to the most complex products occupy the central region of the Product Space (f). Countries in between these two limiting scenarios to occupy products that intermediate the central and peripheral region (e).

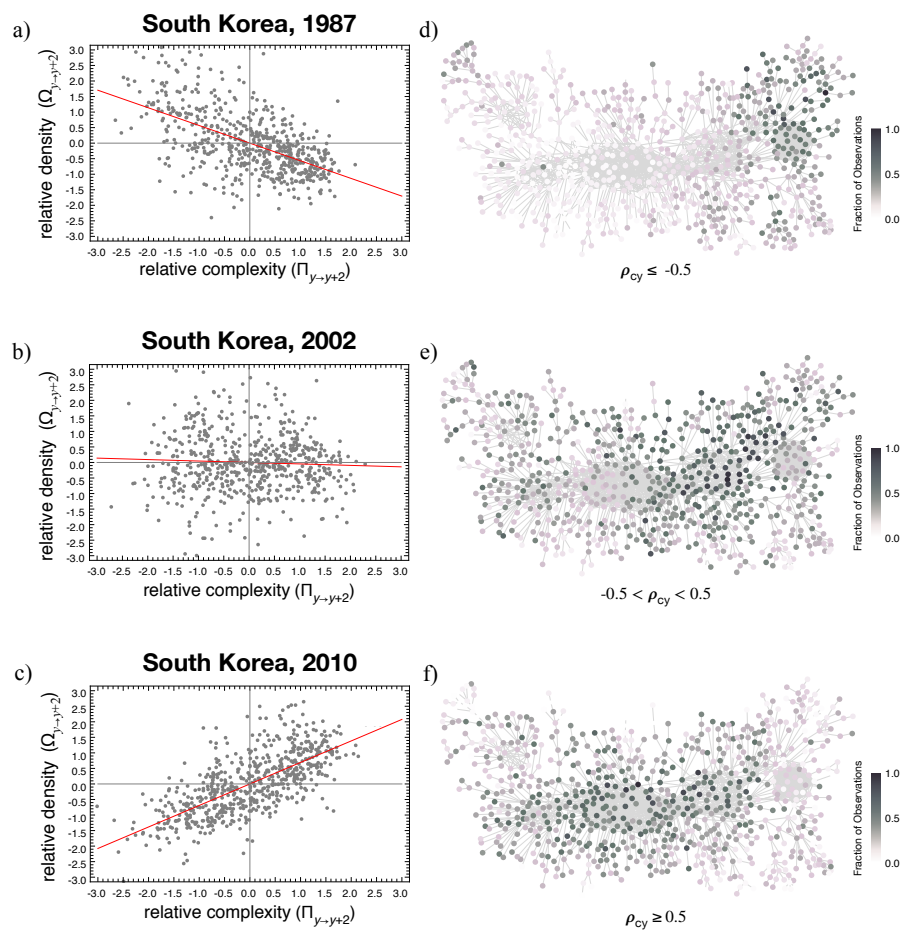

Figure 7 – Examples of the possible correlations between the relative density and complexity of products in the option set of South Korea in the years 1987 (a), 2002 (b) and 2010 (c). The option set

corresponds to all products over which a country has an RCA lower than 1.0. Panels (d), (e) and (f) show the typical distribution of developed products for countries exhibiting the type of correlation exhibited on panels (a), (b) and (c) respectively. Light color indicates that few countries occupy such a product, while a darker color means that a substantial number of countries occupy a product.

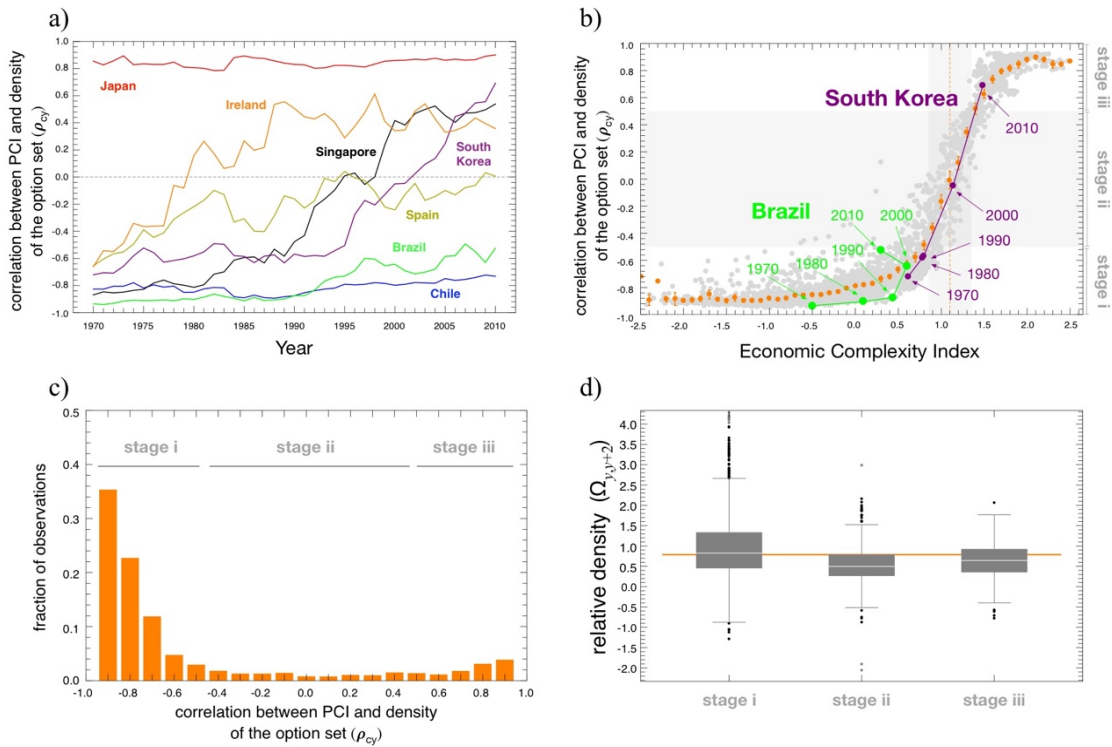

Figure 8 – Panel a) shows the evolution of the correlation between the relative density and relative complexity of the products in the option sets of Japan (dashed red), Ireland (full orange), Singapore (full black), Spain (dashed yellow), South Korea (full purple) and Chile (dashed blue) over time. Panel b) shows the relationship between this correlation and the Economic Complexity index of a country. We use this information to define three stages of development: *i*) when countries are close to low complexity products, ii) when countries' option set do not exhibit a clear correlation; and stage iii) when countries are closer to the most sophisticated products. Panel c) shows a histogram of the observations of countries with different levels of correlations. Panel d) shows the comparison between the distribution in the development direction of densities at different stages of development.

Figure 8a shows the time evolution of the correlation $\rho_{c,y}$ for a number of selected countries, for illustration purposes. Countries, such as Ireland, Singapore and Thailand, underwent a similar transition as South Korea between 1970 and 2010. However, most countries kept a stable profile, like, for instance, Japan and Chile. Some countries fail to fully transit from 'being close to simple products' to 'being close to complex products', such as Brazil and Spain.

Figure 8b shows how the Economic Complexity Index interpolates between countries exhibiting opposite levels of correlation $\rho_{c,y}$; an S-shaped curve can be observed. Highlighted are the evolution of the correlation $\rho_{c,y}$ for the cases of South Korea and Brazil. The S-shaped behavior indicates a rapid transition in diversification opportunities of countries that occurs at precisely the same stage at which countries are more likely to develop towards unrelated varieties. Accordingly, we can divide the diversification opportunities of countries into three stages: i) countries with low complexity: their option set is characterized by a strong negative correlation, bounding these countries to develop towards related and simple products; ii) countries in a rapid transition phase: they benefit from being close to both simple and complex products, which is captured by the lack of correlation between relatedness and complexity of the products in their option set; and iii) countries with highly complex economies: their option set shows a strong positive correlation, making these counties being close to the most complex products.

Figure 8c shows the distribution of countries exhibiting different levels of $\rho_{c,y}$ and illustrates that most countries are close to simple products (stage 1), while some countries—like Japan, Sweden, and the USA— are close to complex products (stage 3), and others–South Korea, Spain and Ireland—underwent a transient scenario and were/are close to both complex and simple products (stage 2).

Figure 8d shows a box plot of the distribution of the development direction of the jumps to either more or less related products $\Omega_{c,y \to y+2}$ at each of the three stages of development. We compared whether the distributions among the different stages are identical through a nonparametric test, namely the two-sample Kolmogorov-Smirnoff test (Wang, Tsang, & Marsaglia, 2003). The results indicate that the three stages are unlikely to follow the same distribution (with a *p-value* lower than 0.001). These results suggest that countries effectively develop towards products with different relative levels of relatedness depending on their stage of development, as the behavior of countries at different stages of development exhibit statistically different patterns.

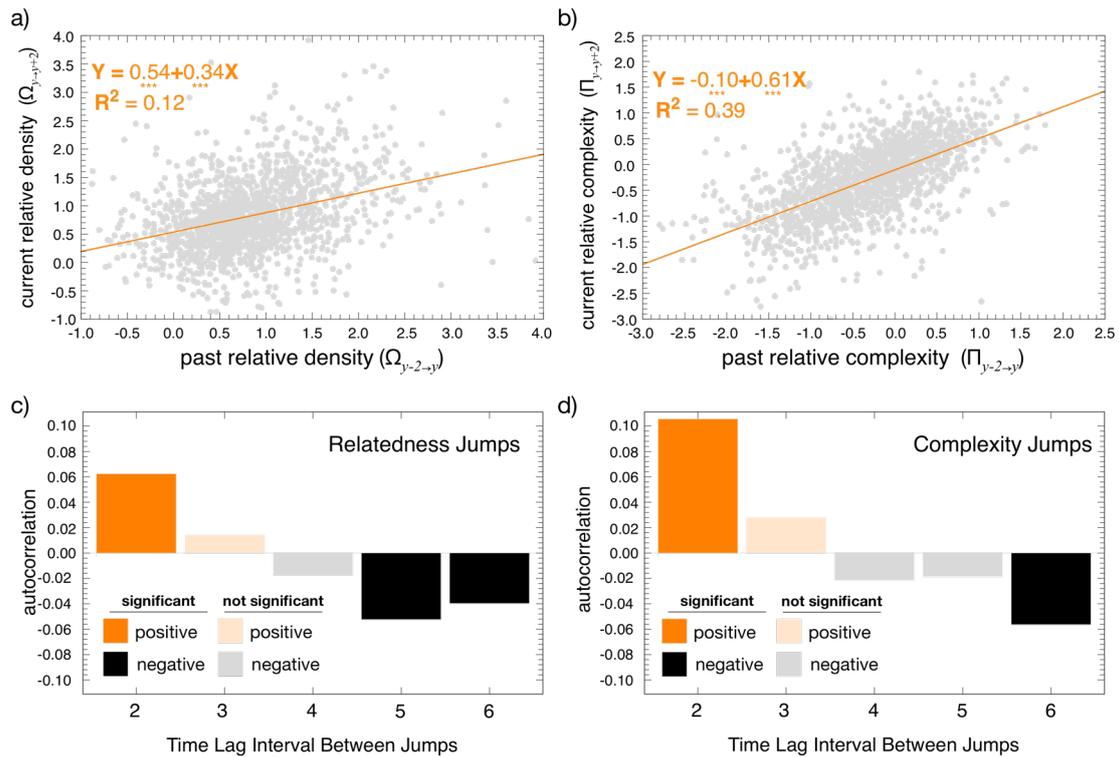

Figure 9 – Correlation between past and current development directions of countries. Panel a and b show the relationship between the past development direction at y-2 and the development direction at y for the respective relative relatedness and relative complexity of newly developed products. Panel c and d show the autocorrelations for different time lags, the first in the case of the relative relatedness and the second for the relative complexity of the newly developed products. Bars correspond to the average autocorrelations over all countries time series; statistical significance was estimated by a t-test to evaluate whether the obtained average was significantly different from zero.

But why do countries develop unrelated activities at an intermediate stage of economic development, when at this point, they have finally access to nearby complex activities?

A possible explanation may come from a natural inertia. Countries reach the intermediate level of development by jumping further, and when they get there, they keep on doing so. Figure 9, shows that this is the case, by exploring the correlation between the relatedness of country's past and future new exports. The positive correlation confirms the idea that countries that are making long or short jumps, continue doing long or short jumps in the subsequent time periods. This correlation, however, is not long-lived, and vanishes after a few years. Thus, unrelated development at an intermediate stage may be the result of a successful path of unrelated development that allowed them to assess such intermediate stage in the first place. It must also be noted that these activities are unrelated for countries at low or intermediate levels of economic development, yet at very high levels of economic development, there might be very few opportunities left for unrelated variety jumps in the product space.

Another explanation could come from strategic considerations. Recently, Alshamsi, Pinheiro & Hidalgo (2018) found that the development strategies that minimize the total time needed to diversify an economy involve entering highly connected activities at an intermediate level of diversification. For countries at an intermediate level of development, this optimal strategy means that jumping to unrelated products may be more beneficial in the long run because of the future diversification opportunities that these open. So even though countries at an intermediate level of development have related and unrelated complex products, entering the more unrelated products may be more beneficial because of the subsequent diversification opportunities these provide. According to the model of Alshamsi, Pinheiro & Hidalgo (2018) only countries at an intermediate level of diversification benefit from entering unrelated activities. For countries with low levels of diversification, this strategy is too risky. For countries with high levels of diversification, strategies focused on unrelated activities are already too late. So, the observation that countries enter more unrelated activities at intermediate levels of diversification would be in line with the predictions of a model in which countries are looking to minimize the total time they need to diversify their economies.

The strategic unrelated jumps at intermediate level of economic development tend to be also in line with the experience of some Asian countries like South Korea and Singapore, where the state has been very active in investing and promoting the development of entirely new sectors either directly, or indirectly through the establishment of a strong education and research infrastructure and key institutions that provided the foundations for making jumps in their industrial evolution (Fagerberg & Srholec, 2008). This happened in combination with attracting foreign direct investments that made countries move in more complex, unrelated products. These multinational companies could rely on complex firm-internal capabilities in their home countries and therefore do well despite being unrelated to the capabilities present in their host countries (Neffke et al., 2018).

## Conclusions

This article attempts to address several key questions in the literature of economic development: To which extent do countries develop towards unrelated activities? Do these events happen at random or are they more likely to happen at a particular stage of development? When countries do so, are they experiencing faster economic growth? And, what factors predict the ability of countries to enter unrelated activities? To answer these questions, we introduced a new methodology that estimates the degree of relatedness and complexity of newly developed products. This approach allows for a comparative analysis of different countries, at different stages of development, and across different time periods.

We have found that, although the trend is for countries to develop towards related activities, the development of unrelated activities was observed in 7.2% of the cases. These events happen more frequently at an intermediate stage of development, which coincides with the time at which economies experience a transforming shift in their productive structure: from being more related to simple products to being more related to complex ones. Finally, and more importantly, we have found that countries that develop towards unrelated activities experience a small but significant boost in economic growth, for instance in a two-year interval, one standard deviation in how unrelated the newly developed products are, leads to a boost of 0.5% in the GDP per capita growth.

## Acknowledgments

The authors acknowledge support from the MIT Media Lab Consortia and from the *Masdar Institute of Technology*. This work was also supported by the *Center for Complex Engineering Systems* (CCES) at *King Abdulaziz City for Science and Technology* (KACST) and the *Massachusetts Institute of Technology* (MIT). The authors are also thankful to Pierre-Alexandre Balland, Mary Kaltenberg, Cristian Jara-Figueroa, Bogang Jung, the remaining Collective Learning Group at the MIT Media Lab for helpful insights and discussions, and for the feedback obtained during the 4th Geography of Innovation Conference in Barcelona, Jan 31-Feb 2, 2018.

# Appendix A – List of countries

The final list of countries includes the following 93 countries: Angola, Argentina, Australia, Austria, Belgium, Bulgaria, Bolivia, Brazil, Canada, Switzerland, Chile, China, Cote d'Ivoire, Cameroon, Republic of the Congo, Colombia, Costa Rica, Cuba, Denmark, Dominican Republic, Algeria, Ecuador, Egypt, Spain, Finland, France, Gabon, United Kingdom, Ghana, Guinea, Greece, Guatemala, Hong Kong, Honduras, Hungary, Indonesia, India, Ireland, Iran, Israel, Italy, Jamaica, Jordan, Japan, Kenya, Cambodia, South Korea, Kuwait, Lebanon, Liberia, Libya, Sri Lanka, Morocco, Mexico, Burma, Mozambique, Mauritius, Malaysia, Nigeria, Nicaragua, Netherlands, Norway, New Zealand, Oman, Pakistan, Panama, Peru, Philippines, Papua New

Guinea, Poland, North Korea, Portugal, Paraguay, Qatar, Romania, Saudi Arabia, Sudan, Singapore, El Salvador, Sweden, Syria, Thailand, Trinidad and Tobago, Tunisia, Turkey, Tanzania, Uruguay, United States, Venezuela, Vietnam, South Africa, Zambia and Zimbabwe.

# Appendix B – Data Analysis

In this appendix, we extend the analysis conducted in the manuscript. We provide a description of the data and further statistical analysis of the country level features used in the regression models shown in Table 2 of the main text. We start by showing the correlation between the different independent variables. These correspond to both variables proposed in the manuscript, measuring the level of relative relatedness ($\Omega_{c,y \rightarrow y+2}$) and complexity ($\Pi_{c,y \rightarrow y+2}$) of newly developed products, as well as country-specific features. The results are shown in Figure B1.

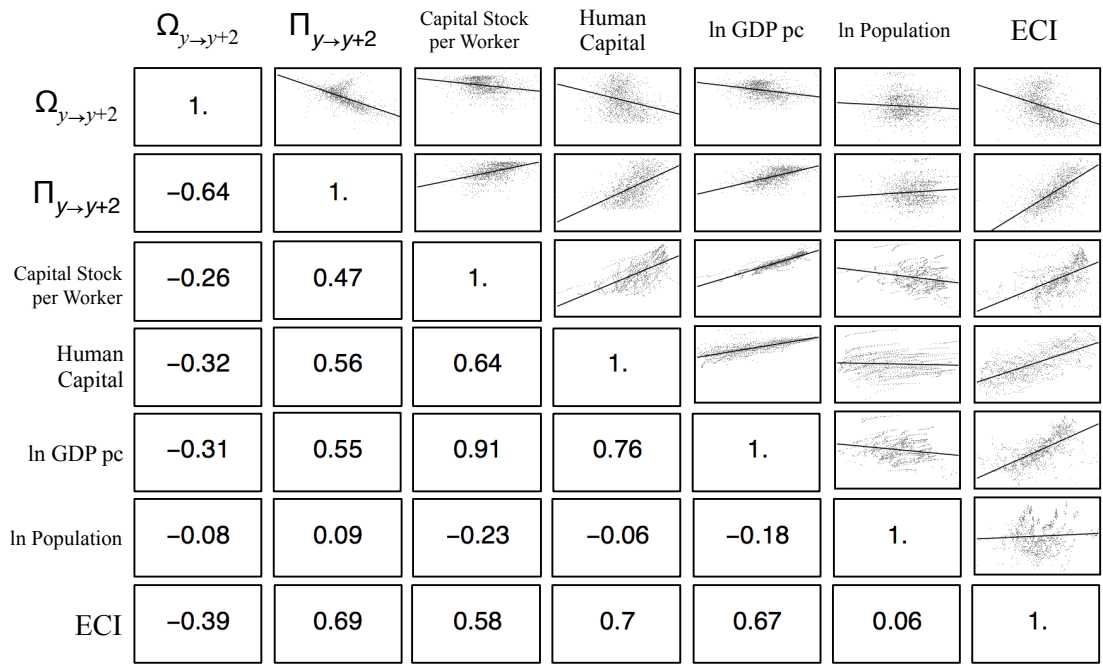

Figure B1 – Correlation Matrix between the independent variables and the proposed variables, namely the relative relatedness of newly developed products ($\Omega_{c,y \rightarrow y+2}$), relative complexity of newly developed products ($\Pi_{c,y \rightarrow y+2}$), Capital Stock per Worker, Human Capital Index, ln GDP per capita, ln Population, and Economic Complexity Index.

Next, we analyze the level of multicollinearity by computing the Variance Inflation Factor (VIF) associated with each variable in the regressions presented in Table 2.

The VIF quantifies the degree of the multicollinearity among the different variables; the results are reported in Table B1.

Table B1 – Variance Inflation factors associated with the variables used in the Regression models depicted on Table 2.

|  | (1) | (2) | (3) | (4) | (5) | (6) | (7) |
|---|---|---|---|---|---|---|---|
| $\Omega_{c,y \to y+2}$ | 2.48 |  | 4.24 | 2.98 |  |  | 4.29 |
| $\Pi_{c,y \to y+2}$ |  | 1.24 | 2.11 |  |  | 2.50 | 3.61 |
| Initial ECI |  |  |  | 2.86 | 2.96 | 3.50 | 3.52 |
| Initial ln GDP$_{pc}$ |  |  |  | 569.31 | 569.31 | 573.27 | 574.22 |
| Initial ln Population |  |  |  | 5.81 | 5.85 | 5.88 | 5.89 |
| Initial Human Capital |  |  |  | 45.53 | 45.64 | 45.58 | 45.65 |
| Capital Stock per Worker |  |  |  | 466.13 | 466.27 | 466.13 | 466.40 |

As shown in Figure B1, and reinforced by the VIF analysis, the new variables relative density and relative complexity show a small degree of multicollinearity with the remaining independent variables, and among themselves. The large multicollinearity observed in GDP per capita and Human Capital was expected, as the ECI has been shown, previously, to be highly correlated with these two variables.

Next, we show that the interaction term between *relative relatedness* and *complexity* is not statistically significant. In Table B2, we show models that regress the GDP per capita growth over two-year intervals, after entering new activities, as a function of the relative relatedness and complexity of the newly developed products and the interaction term between them. The interaction term is statistically not significant. Moreover, we note that its inclusion in the model does not increase the explained level of variance, see model (4) when compared with models (1) and (3). This shows that the interaction term does not improve the quality of the regression. Further inspection can be done by performing an F-test between models (3) and (4), which results in the non-significant F-statistics of $0.2422$ ($p$-value $= 0.63$). For these reasons, we did not include the interaction term in the models depicted in Table 2.

Table B2 – Linear models regressing the GDP per capita annualized growth over two years, after entering new products, as a function of the relative relatedness ($\Omega_{c,y \to y+2}$) and complexity ($\Pi_{c,y \to y+2}$) of the newly developed products.

|  | GDP per capita annualized growth ($g_{c,y \to y+2}$) | | | |
| --- | --- | --- | --- | --- |
|  | (1) | (2) | (3) | (4) |
| $\Omega_{c,y \to y+2}$ | -0.005*** |  | -0.005*** | -0.006*** |
|  | (0.001) |  | (0.002) | (0.002) |
| $\Pi_{c,y \to y+2}$ |  | 0.003** | -0.00003 | 0.001 |
|  |  | (0.001) | (0.002) | (0.002) |
| $\Omega_{c,y \to y+2} \times \Pi_{c,y \to y+2}$ |  |  |  | -0.001 |
|  |  |  |  | (0.001) |
| Constant | 0.041*** | 0.038*** | 0.041*** | 0.041*** |
|  | (0.004) | (0.004) | (0.004) | (0.004) |
| Observations | 1,511 | 1,511 | 1,511 | 1,511 |
| $R^2$ | 0.079 | 0.073 | 0.079 | 0.079 |
| F Statistic | 6.386*** | 5.871*** | 6.078*** | 5.810*** |

Notes: *p<0.1; **p<0.05; ***p<0.01
All models control for year fixed effect
Observations collected between 1970 and 2010 non-overlapping 2 years steps

Next, we show the consistency of models (1) and (5) from Table 2, in particular how the coefficient associated with $\Omega_{c,y \to y+2}$ behaves under variations of the time interval used to identify newly developed products (δ). We see that the impact of varying these two parameters is negligible both regarding the magnitude and value of the coefficient: it remains negative in all but one case and with values around -0.005. The level of statistical significance, when controlled by other country-specific factors, depends on the choice of parameters.

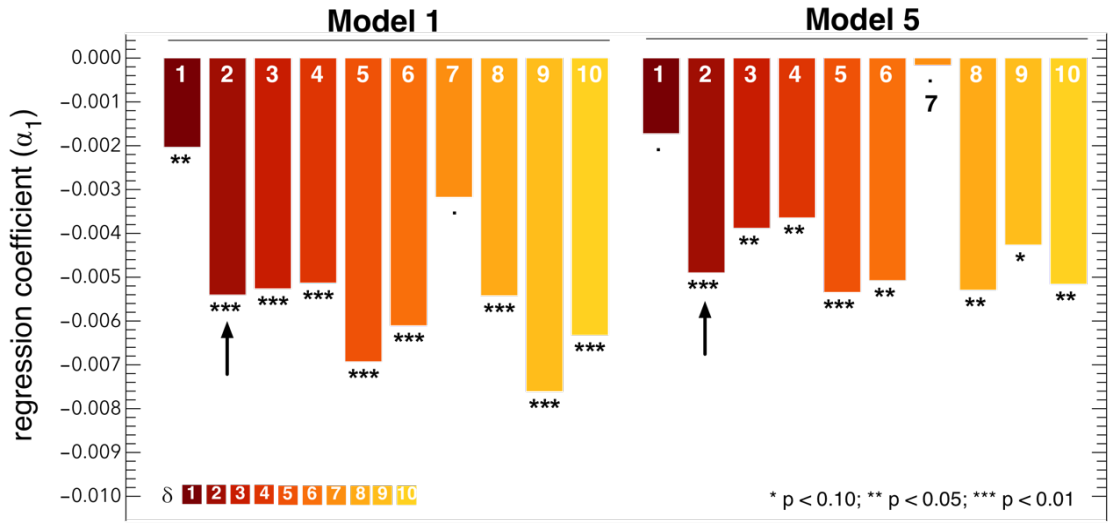

Figure B2 – Bar plot showing the coefficient associated with $\Omega_{c,y\to y+\delta}$ as in regression models (1) and (5) of Regression Table 2 under different conditions. Each bar depicts a different choice of time interval ($\delta$) to estimate the GDP per capita growth and the new products entered by a country. The level of statistical significance of each coefficient is also depicted bellow each bar. Arrows point towards the choice of $\delta = 2$ depicted in Table 2 and discussed in the main manuscript.

Finally, we show a summary description of the variables of interest used in the analysis conducted. In Table B3 and B4, each row shows, ordered by year, the number of countries with observable jumps ($N$), average size of the option set ($O$), average size of the product basket ($P$), total number of observed jumps ($N_j$) average deviations in the relatedness of developed products ($\Omega_{c,y\to y+2}$), average deviations in the complexity of developed products ($\Pi_{c,y\to y+2}$ ), average ECI, average correlation between the relatedness and complexity of the option set ($\rho$), Log of the average GDP *per capita*, Log of the average Population size, the Human Capital Index, and the Log of the Capital Stock per employed worker.

| Year | N | O | P | $N_j$ | $\Omega_{c,y \to y+2}$ | $\Pi_{c,y \to y+2}$ | ECI | $\rho$ | Log $GDP_{pc}$ | Log Pop | HC | Stock |
|---|---|---|---|---|---|---|---|---|---|---|---|---|
| 1970 | 84.00 | 483.65 | 87.35 | 448.00 | 0.86 | -0.39 | 0.11 | -0.59 | 4.77 | 1.00 | 1.81 | 10.67 |
| 1971 | 82.00 | 487.96 | 90.04 | 474.00 | 0.83 | -0.38 | 0.13 | -0.59 | 4.79 | 1.01 | 1.83 | 10.70 |
| 1972 | 84.00 | 486.69 | 91.31 | 558.00 | 0.74 | -0.34 | 0.16 | -0.58 | 4.80 | 0.98 | 1.84 | 10.80 |
| 1973 | 82.00 | 484.72 | 93.28 | 525.00 | 0.77 | -0.34 | 0.18 | -0.58 | 4.83 | 1.00 | 1.86 | 10.82 |
| 1974 | 82.00 | 510.04 | 101.96 | 460.00 | 0.88 | -0.27 | 0.17 | -0.55 | 4.89 | 1.04 | 1.92 | 10.83 |
| 1975 | 78.00 | 517.91 | 108.09 | 464.00 | 0.78 | -0.25 | 0.21 | -0.54 | 4.89 | 1.05 | 1.91 | 10.82 |
| 1976 | 78.00 | 579.88 | 123.12 | 416.00 | 0.89 | -0.38 | 0.15 | -0.50 | 4.91 | 1.04 | 1.93 | 10.92 |
| 1977 | 83.00 | 587.54 | 118.46 | 397.00 | 0.85 | -0.33 | 0.10 | -0.49 | 4.94 | 1.05 | 1.95 | 10.99 |
| 1978 | 82.00 | 607.45 | 119.55 | 413.00 | 0.82 | -0.25 | 0.20 | -0.49 | 4.97 | 1.07 | 2.00 | 11.00 |
| 1979 | 74.00 | 575.64 | 122.36 | 374.00 | 0.80 | -0.30 | 0.27 | -0.50 | 4.99 | 1.11 | 2.03 | 11.00 |
| 1980 | 68.00 | 577.43 | 128.57 | 249.00 | 0.96 | -0.34 | 0.32 | -0.53 | 4.98 | 1.10 | 2.00 | 11.02 |
| 1981 | 70.00 | 588.96 | 127.04 | 266.00 | 1.02 | -0.54 | 0.23 | -0.51 | 4.98 | 1.08 | 2.02 | 11.08 |
| 1982 | 86.00 | 593.98 | 112.02 | 663.00 | 0.80 | -0.63 | 0.10 | -0.51 | 4.98 | 1.08 | 2.02 | 11.09 |
| 1983 | 86.00 | 593.66 | 109.34 | 749.00 | 0.77 | -0.57 | 0.11 | -0.52 | 5.00 | 1.11 | 2.05 | 11.11 |
| 1984 | 89.00 | 720.97 | 128.03 | 438.00 | 1.02 | -0.58 | 0.08 | -0.59 | 5.01 | 1.11 | 2.06 | 11.10 |
| 1985 | 89.00 | 715.72 | 122.28 | 455.00 | 1.03 | -0.55 | 0.05 | -0.61 | 5.01 | 1.12 | 2.08 | 11.10 |
| 1986 | 89.00 | 662.22 | 110.78 | 499.00 | 1.02 | -0.53 | 0.03 | -0.60 | 5.05 | 1.14 | 2.11 | 11.12 |
| 1987 | 82.00 | 654.73 | 119.27 | 513.00 | 0.97 | -0.53 | 0.08 | -0.57 | 5.10 | 1.18 | 2.17 | 11.12 |
| 1988 | 83.00 | 649.80 | 126.20 | 588.00 | 0.95 | -0.41 | 0.12 | -0.56 | 5.10 | 1.18 | 2.19 | 11.14 |
| 1989 | 83.00 | 648.19 | 127.81 | 611.00 | 1.04 | -0.45 | 0.16 | -0.56 | 5.11 | 1.18 | 2.21 | 11.15 |

Table B3 – Descriptive summary of the variables between 1970 to 1989.

| Year | N | O | P | $N_j$ | $\Omega_{c,y,y+2}$ | $\Pi_{c,y,y+2}$ | ECI | $\rho$ | Log GDP$_{pc}$ | Log Pop | HC | Stock |
|---|---|---|---|---|---|---|---|---|---|---|---|---|
| 1990 | 88.00 | 652.67 | 123.33 | 722.00 | 1.09 | -0.51 | 0.12 | -0.54 | 5.13 | 1.19 | 2.23 | 11.16 |
| 1991 | 86.00 | 647.00 | 128.00 | 741.00 | 0.98 | -0.43 | 0.13 | -0.55 | 5.13 | 1.20 | 2.23 | 11.15 |
| 1992 | 86.00 | 644.63 | 130.37 | 737.00 | 0.93 | -0.31 | 0.10 | -0.55 | 5.13 | 1.19 | 2.25 | 11.19 |
| 1993 | 89.00 | 644.87 | 129.13 | 696.00 | 0.88 | -0.28 | 0.04 | -0.54 | 5.14 | 1.18 | 2.28 | 11.21 |
| 1994 | 88.00 | 640.09 | 131.91 | 638.00 | 0.85 | -0.30 | 0.05 | -0.53 | 5.15 | 1.19 | 2.29 | 11.22 |
| 1995 | 89.00 | 639.33 | 130.67 | 745.00 | 0.89 | -0.34 | 0.04 | -0.53 | 5.18 | 1.21 | 2.32 | 11.22 |
| 1996 | 90.00 | 639.52 | 130.48 | 666.00 | 0.85 | -0.36 | -0.01 | -0.53 | 5.19 | 1.20 | 2.33 | 11.25 |
| 1997 | 88.00 | 636.94 | 133.06 | 501.00 | 0.86 | -0.33 | -0.02 | -0.53 | 5.20 | 1.22 | 2.33 | 11.24 |
| 1998 | 88.00 | 638.65 | 131.35 | 474.00 | 0.86 | -0.25 | -0.07 | -0.53 | 5.22 | 1.24 | 2.37 | 11.24 |
| 1999 | 83.00 | 631.36 | 138.64 | 523.00 | 0.70 | -0.19 | 0.02 | -0.52 | 5.22 | 1.23 | 2.40 | 11.26 |
| 2000 | 84.00 | 629.20 | 141.80 | 583.00 | 0.77 | -0.23 | 0.08 | -0.51 | 5.22 | 1.23 | 2.43 | 11.24 |
| 2001 | 85.00 | 628.67 | 142.33 | 566.00 | 0.83 | -0.30 | 0.10 | -0.51 | 5.24 | 1.22 | 2.47 | 11.31 |
| 2002 | 83.00 | 627.49 | 143.51 | 584.00 | 0.79 | -0.23 | 0.10 | -0.50 | 5.26 | 1.25 | 2.51 | 11.32 |
| 2003 | 85.00 | 628.44 | 141.56 | 589.00 | 0.76 | -0.23 | 0.12 | -0.48 | 5.29 | 1.26 | 2.54 | 11.33 |
| 2004 | 79.00 | 619.44 | 150.56 | 523.00 | 0.77 | -0.22 | 0.16 | -0.47 | 5.32 | 1.27 | 2.55 | 11.36 |
| 2005 | 76.00 | 615.88 | 155.12 | 481.00 | 0.70 | -0.15 | 0.18 | -0.43 | 5.36 | 1.29 | 2.57 | 11.44 |
| 2006 | 81.00 | 622.21 | 148.79 | 494.00 | 0.83 | -0.20 | 0.17 | -0.43 | 5.40 | 1.30 | 2.59 | 11.47 |
| 2007 | 72.00 | 610.50 | 160.50 | 416.00 | 0.65 | -0.17 | 0.26 | -0.43 | 5.42 | 1.32 | 2.58 | 11.47 |
| 2008 | 73.00 | 614.56 | 155.44 | 428.00 | 0.69 | -0.07 | 0.22 | -0.42 | 5.44 | 1.31 | 2.61 | 11.52 |
| 2009 | 81.00 | 622.00 | 147.00 | 967.00 | 0.62 | -0.15 | 0.11 | -0.42 | 5.44 | 1.31 | 2.62 | 11.56 |
| 2010 | 82.00 | 620.38 | 145.62 | 1369.00 | 0.61 | -0.12 | 0.11 | -0.41 | 5.46 | 1.31 | 2.64 | 11.59 |

Table B4 – Descriptive summary of the variables between 1990 to 2010.